\documentclass[12pt]{iopart}
\usepackage{epsfig}    
\usepackage{cite}      
\usepackage{subfigure}

\begin{document}

\letter{Ionization suppression of excited atomic states beyond the
  stabilization regime}

\author{Andreas Staudt \dag\ddag, Christoph H Keitel \dag\ddag \ 
  and John S Briggs \ddag}
\address{\dag Max-Planck-Institut f\"ur Kernphysik,
  Saupfercheckweg 1,
  D-69117 Heidelberg, Germany
}
\address{\ddag Theoretische Quantendynamik,
  Physikalisches Institut,
  Universit\"at Freiburg,
  Hermann-Herder-Stra{\ss}e 3,
  D-79104 Freiburg, Germany
}


\eads{\mailto{astaudt@physik.uni-freiburg.de},
  \mailto{keitel@mpi-hd.mpg.de} and \mailto{briggs@physik.uni-freiburg.de}\\
  {\tt http://www.mpi-hd.mpg.de/keitel/}}

\begin{abstract}
  A two-dimensional model atom is employed to study the ionization
  behavior of initially excited atomic states in highly-frequent intense laser
  pulses beyond the dipole approximation.
  An additional regime of ionization suppression is found at laser intensities
  where the stabilization effect is expected to break down.
  The appearance of this effect is due to a strong coupling of the
  initial wave function to the ground state of the cycle-averaged
  space-translated ionic potential, followed by a subsequent population
  transfer to the ground state during the laser pulse turn-off.
  Non-dipole effects are found to increase the overall ionization probabilities,
  but not to suppress or alter this effect.
\end{abstract}


In recent years, the rapid progress in laser technology has enabled the
research on highly nonperturbative phenomena of atoms in intense laser fields,
such as Above Threshold Ionization (ATI)
or High Harmonic Generation (HHG) \cite{Burnett_93}.
With the advent of Free Electron Lasers (FEL) \cite{Materlik_01},
light sources will soon be available to generate photons
whose energy $\hbar \omega$ may equal or even exceed the binding energies
of ground state atoms.
At such high frequencies, the atom may stabilize against
ionization \cite{Gavrila_92},
such that the ionization probability must not necessarily rise with
the laser intensity,
but it may decrease even though the intensity is increased.
The stabilization effect and dynamic ionization suppression have been
extensively studied in one-electron atoms \cite{Gavrila_84},
and also two-electron systems have been considered \cite{Mittleman_90}.
Stabilization of initially excited atomic states has been of interest
since they provide a means to fulfill the condition of high
laser frequencies, that is, within these systems the energy
of a single photon exceeds the electronic binding energy.
Theoretical investigations have been carried out on the ionization
of 2s and 2p states of hydrogen \cite{Doerr_91},
while the existence of the stabilization effect has been experimentally
verified on Rydberg states \cite{Boer_94}.

In this letter, we study the ionization dynamics of an initially excited model
atom subjected to short, highly intense laser pulses
whose frequency exceeds the atomic ground state binding energy.
Next to the stabilization effect of the atom, which is expected to occur
at the laser parameters employed in our calculations,
a second stage of ionization suppression at higher field strengths
can be observed,
which becomes more pronounced for longer laser pulses.
This effect can be understood by means of a quasi-static picture,
where the initially excited atomic state couples strongly to the
ground state of the time-averaged Kramers-Henneberger potential.
Furthermore, an alternative picture is employed where the suppression effect
is explained in terms of the evolution of the central regions with small
probabilities of the excited state wave packet.

\begin{figure}[t]
  \subfigure{
    \epsfig{file=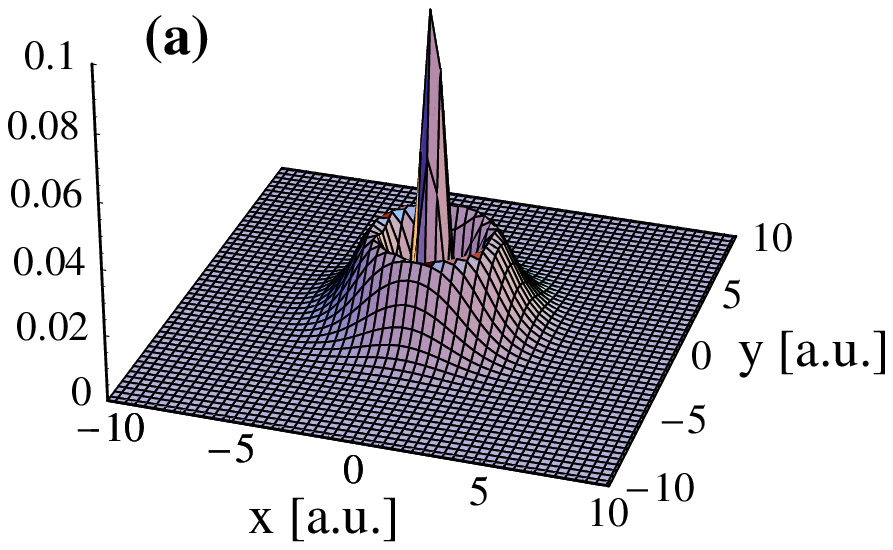, width=0.475\textwidth, clip=true}
    }
    \subfigure{
    \epsfig{file=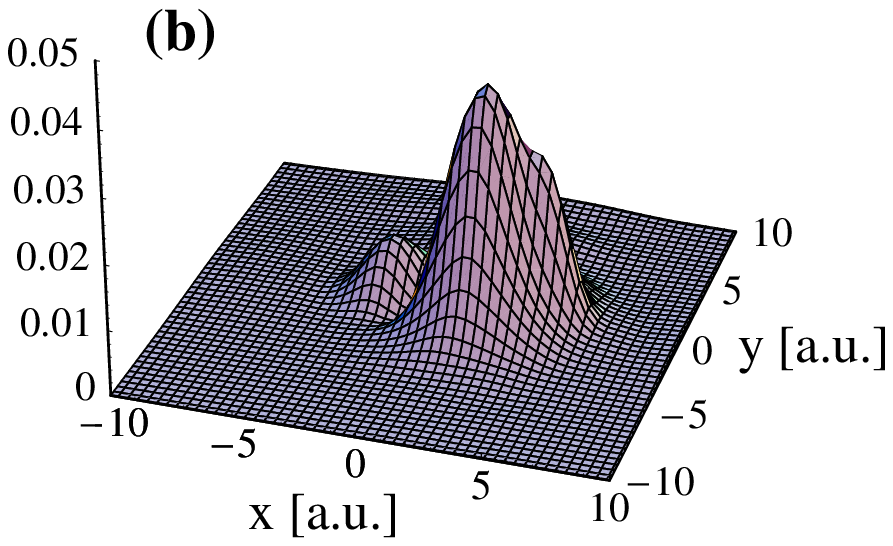, width=0.475\textwidth, clip=true}
    }
  \caption{Probability densities of the initial (2s) state (a)
    and the same wave function after eight optical cycles interaction
    with a laser pulse of frequency $\omega=3$ a.u. and peak field strength
    $E_0=40$ a.u. (b).
    The x-axis denotes the laser polarization direction,
    while the pulse travels in the positive y-direction.}
  \label{He3_Dens}
\end{figure}

A two-dimensional model atom \cite{Aldana_00, Kylstra_00} has been employed
in our calculations,
which allows for the inclusion of non-dipole effects of the laser.
The electronic wave packet is restricted to the plane spanned by the
laser polarization and propagation direction.
To account for the reduced dimensionality of the system, the atomic potential
is described by a two-dimensional soft-core potential \cite{Javanainen_88}
(atomic units are used)
\begin{equation}\label{Soft_Core}
  V_{\rm SC}({\bf r}) =
  - \frac{3.28}{\sqrt{{\bf r}^2 + 1}}.
\end{equation}
The soft-core parameters in (\ref{Soft_Core}) have been chosen to
reproduce the ground state energy of the helium ion.
Then, the excited state 2s $\Phi_{\rm 2s}$ has the binding energy
$E_B=0.832$ a.u.,
and will be used as initial state with probability density displayed
in Fig. \ref{He3_Dens} (a).
The wave function then is propagated with respect to time using the
split-operator method \cite{Aldana_00, Hu_01}.
The interaction with the linearly polarized laser pulse is described by
the vector potential ${\bf A}({\bf r}, t)$ which,
due to its spatial dependence, accounts for the retardation of the laser.
Both ${\bf A}$ as well as the electric field along the laser polarization
direction are continuous and chosen to have a trapezoidal envelope,
while it is ensured that no purely field-induced net momentum is transferred to
the electron after the interaction with the pulse,
i.e. the time integral over the pulse vanishes \cite{Geltman_99}.
In the following, we restrict the duration of the turn-on and turn-off stages
of the pulses to one optical cycle each,
while the length of the intermediate stage at constant amplitude is varied.
High laser frequencies of $\omega=3$ a.u. are considered,
such that the energy of one photon exceeds even the ionic ground state energy,
and stabilization may occur.
The crucial quantity we wish to observe after the atom has been exposed
to a laser pulse is the ionization probability.
Since the calculations of all the bound as well as the free states
of the atom is far too complex to obtain a good enough resolution,
we employ an alternative scheme to define the ionization probability.
At every point of the numerical grid we calculate the sum of potential
and kinetic energy.
If the total energy is positive, the portion of the electronic wave packet
stored on the respective point of the grid is expected to leave the vicinity
of the nucleus and finally become ionized.
The validity of this method has been verified by explicitly projecting out
a series of previously calculated bound states before calculating
the ionization probability with the method described above,
which quantitatively yields the same results.

\begin{figure}[t]
  \epsfig{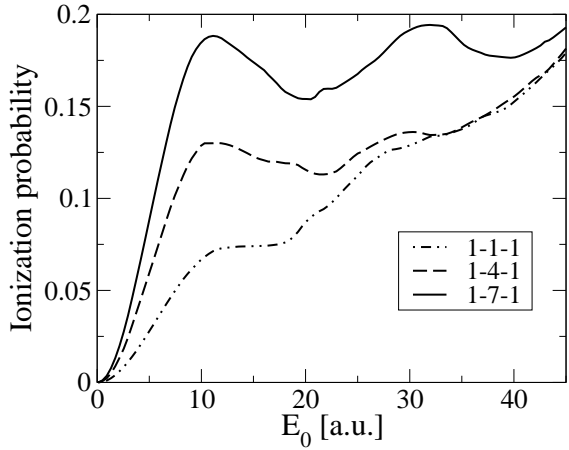}
  \caption{Ionization probabilities of the first excited state
    as a function of the peak electric field strength $E_0$
    at a frequency $\omega=3$ a.u. for three different pulse lengths.
    Trapezoidal pulse envelopes have been employed in the calculations,
    with one optical cycle linear turn-on and turn-off in each case,
    and central portions of length one, four and seven cycles,
    respectively.}
  \label{He3_1X1}
\end{figure}

In Fig. \ref{He3_1X1} the ionization probabilities of the excited model atom
are displayed for three different pulse lengths,
where the number of turn-on and turn-off cycles has been kept fixed at one each,
while the length of the centre portions of the pulse at constant
peak amplitude $E_0$ has been varied.
For all three pulses strong deviations of the ionization probabilities
from a monotonic increase with field strength can be observed for
$E_0 > 10$ a.u., and stabilization sets in.
While for the shortest pulse of three cycles total duration mere saturation
of the ionization probabilities can be observed,
for longer pulse durations a decrease clearly sets in.
At field strengths $E_0 > 20$ a.u. the atomic stabilization breaks down,
and the ionization probabilities again tend to increase with the laser
intensity \cite{Gajda_92}.
However, for total pulse durations of more than than six cycles,
an additional decrease in the ionization probabilities is observable
for field strengths of about $E_0=35$ a.u.,
which is not present in the case of the ground state of the atomic system.
Our calculations have shown that this effect persists for even longer
pulse durations,
while it gradually diminishes when the length of the turn-on and turn-off
of the laser pulses is increased.
To understand the origins of this effect, we switch into the eigensystem
of the unperturbed electron in the laser field,
described quantum-mechanically by the Kramers-Henneberger (KH)
transformation \cite{Kramers_65},
in which the electron at rest is subject to the oscillating atomic
potential.
This frame has been adopted successfully to explain the stabilization effect,
where at high laser frequencies the electronic wave packet tends to occupy
the ground state $\Phi_{\rm KH}$ of the time-averaged KH potential,
$\overline{V}_{\rm KH}({\bf r})$.
This is a double-well potential with the two wells separated by twice
the quiver amplitude $\alpha_0 = E_0 / \omega^2$
along the laser polarization direction ${\bf \hat{\epsilon}}$,
and can be calculated by time-averaging over one laser period
$T=2 \pi /\omega$ via
\begin{equation}\label{KH_Potential}
  \overline{V}_{\rm KH}({\bf r}) =
  \frac{1}{T} \int_0^T V_{\rm SC}({\bf r} 
  + \alpha_0 \cos\left( \omega t \right) {\bf \hat{\epsilon}})
  \ {\mathrm d} t \, .
\end{equation}
This explains why the stabilization effect at field strengths
$E_0 \approx 20$ a.u., Fig. \ref{He3_1X1},
becomes more pronounced with the pulse duration,
since with longer pulses the electron evolves more effectively into
the ground state of the averaged KH potential.
For the second ionization suppression effect at about $E_0=40$ a.u.
a different mechanism comes into play.
The initially excited population gradually evolves into the ground state
of $\overline{V}_{\rm KH}$,
while being transferred into the field-free atomic ground state
during the rapid turn-off of the laser pulse.

\begin{figure}[t]
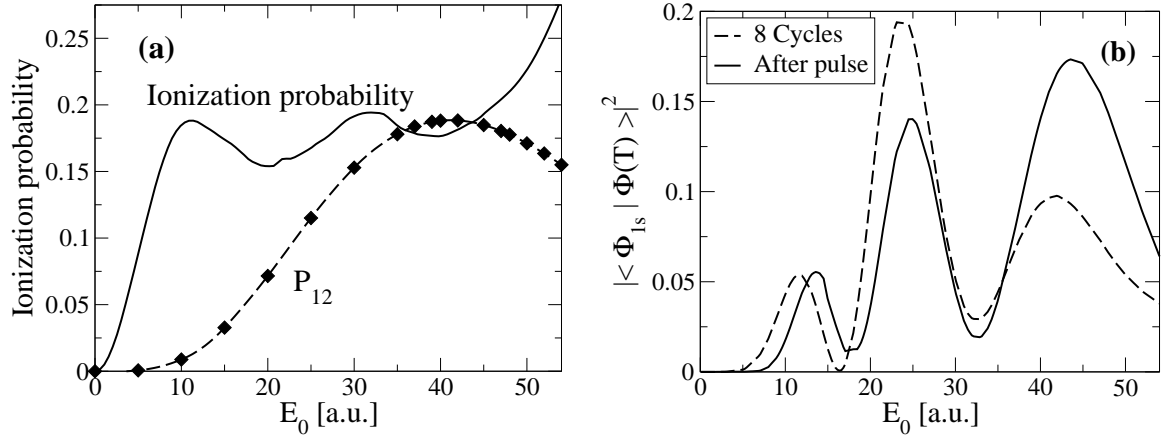

  \subfigure{
    \epsfig{file=staudtfig3a.eps, width=0.475\textwidth, clip=true}
  }
  \subfigure{
    \epsfig{file=staudtfig3b.eps, width=0.475\textwidth, clip=true}
  }
  \caption{(a) Ionization probability of the initially excited state
    after the interaction with a laser pulse of nine optical cycles
    total duration at frequency $\omega=3$ a.u..
    The chained curve denotes the expectation value $P_{12}$,
    as defined by Eq. (\ref{Transition_Operator}).
    The maximal field strength $E_0=54$ a.u. for which results are shown
    implies an excursion amplitude $\alpha_0=6$ a.u..
    (b) Occupation probabilities of the ground state after eight
    optical cycles interaction and after the laser pulse has been ramped down.
  }
  \label{KH_Overlap}
\end{figure}

In Fig. \ref{KH_Overlap} (a), in addition to the ionization probabilities
after interaction with a 1-7-1 pulse the expectation value of the overlap
operator $O=\left| \Phi_{\rm 1s}\right> \left< \Phi_{\rm 2s} \right|$
between the initially excited 2s state and the 1s state
in the state $\Phi_{\rm KH}$ is displayed.
This expectation value is given by
\begin{equation}\label{Transition_Operator}
  P_{12}=\left| \left< \Phi_{\rm KH} | O | \Phi_{\rm KH}\right> \right|^2 
  = \left| \left< \Phi_{\rm KH} | \Phi_{\rm 1s}\right>
    \left< \Phi_{\rm 2s} | \Phi_{\rm KH}\right> \right|^2.
\end{equation}
In these terms, the ground state in the potential $\overline{V}_{\rm KH}$
mediates the laser-assisted population transfer from the initial to the
field-free ground state.
Even though this representation by overlap matrix elements provides
only a rough estimate of the underlying mechanism,
since the time-dependence of the process is neglected,
the position of the maximum of this quantity coincides well with the
field strength at which the ionization suppression effect occurs.
This view is further corroborated when considering the overlap between
the time-dependent wave function and the field-free ground state,
Fig. \ref{KH_Overlap} (b).
One clearly sees that between the beginning of the turn-off of the laser pulse
at eight cycles and the end of the pulse,
at field strengths about $E_0=40$ a.u.,
the population transfer to the ground state is enhanced,
leading to the effective suppression of ionization in this intensity regime.

\begin{figure}[t]
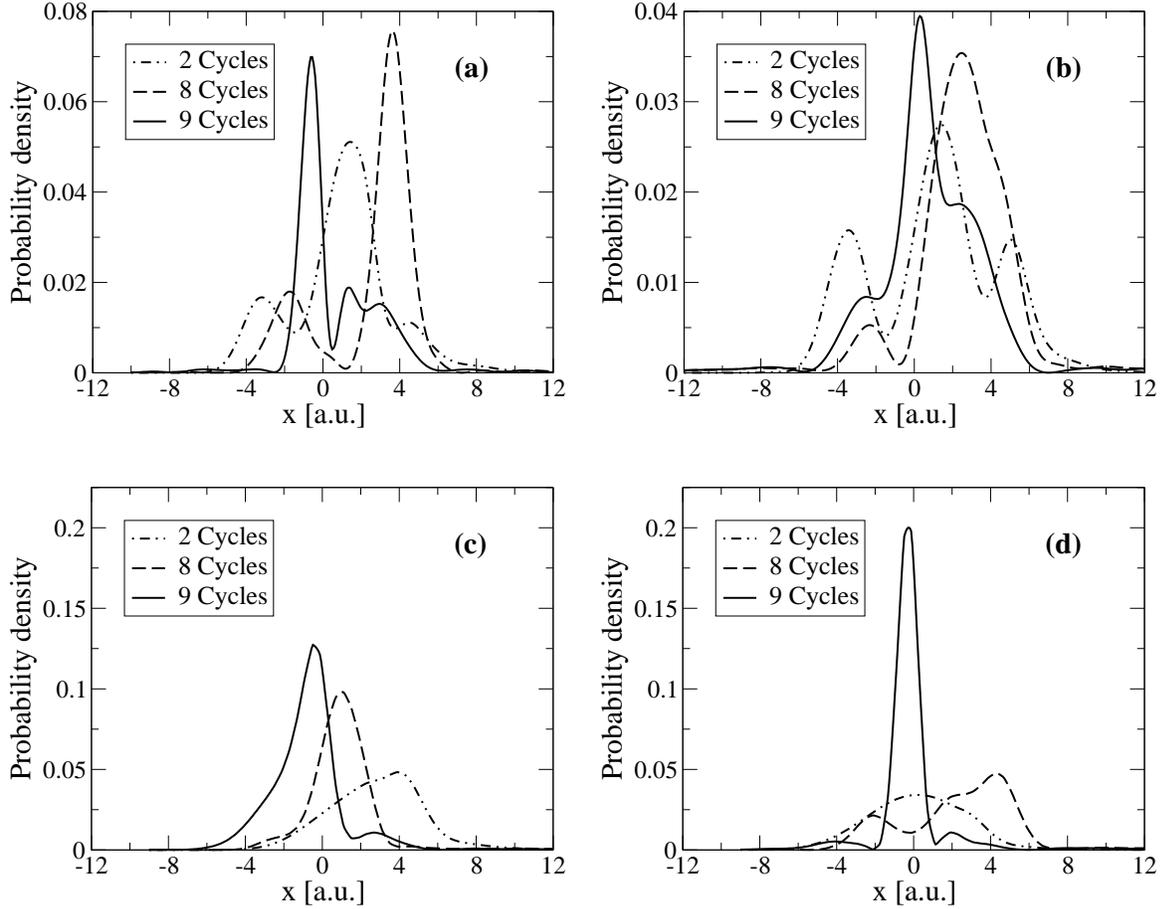

  \subfigure{
    \epsfig{file=staudtfig4a.eps, width=0.475\textwidth, clip=true}
  }
  \subfigure{
    \epsfig{file=staudtfig4b.eps, width=0.475\textwidth, clip=true}
  }\\
  \subfigure{
    \epsfig{file=staudtfig4c.eps, width=0.475\textwidth, clip=true}
  }
  \subfigure{
    \epsfig{file=staudtfig4d.eps, width=0.475\textwidth, clip=true}
  }
  \caption{Density profiles along the laser polarization axis $\hat{\epsilon}$
    during the interaction with a nine-cycle pulse of frequency
    $\omega=3$ a.u. and peak field strengths $E_0=30$ a.u.
    (left column) and $E_0=40$ a.u. (right column), respectively.
    In (a) and (b), the density profiles are shown for the excited (2s)
    initial state,
    while in (c) and (d) the ground (1s) state has been chosen
    as initial state.
    Notice the change in scale between (a) and (b).
  }
  \label{Density_Profiles}
\end{figure}

The appearance of the ionization suppression effect can also be understood
by considering the evolution of the laser-driven wave packets in time.
Displayed in Fig. \ref{Density_Profiles} are density profiles
of the wave functions along the laser polarization axis,
i.e. for $y=0$.
When driven by the laser pulse, the wave packets lose their initial shape
exhibiting three distinct maxima separated by nodes at $x=\pm 1.2$ a.u.,
and evolve to functions showing a doubly-peaked structure,
Fig. \ref{Density_Profiles} (a), (b).
At $E_0=30$ a.u., the wave function at $t=8$ cycles exhibits substantial
probability density in the area around the nucleus.
During the turn-off of the laser pulse,
the density profile evolves to a shape showing a minimum in the vicinity
of the origin.
This is in contrast to the wave function during the turn-off of the laser pulse
in the case $E_0=40$ a.u., Fig. \ref{Density_Profiles} (b).
At $t=8$ cycles, the position of the wave function minimum is closer
to the origin than for $E_0=30$ a.u.,
and the absolute value of the density probability is smaller than in
Fig. \ref{Density_Profiles} (a),
therefore only a small portion of the electronic wave packet is
located in the vicinity of the nucleus.
Consequently, the ionization cross section is reduced,
such that population is transferred to the ground state of the potential
during the ramping off of the pulse.
This can also be deduced from  Fig. \ref{Density_Profiles} (b) at $t=9$ cycles,
where the density shows a peak at the origin, similar to the 1s state.
The ionization suppression effectively works for laser pulses
longer than six optical cycles duration, as can be seen
from Fig. \ref{Density_Profiles} (a) and (b),
where suppression is seen not to occur for $t=2$ cycles,
since the electronic wave packet first has to evolve into the highly distorted
shape with distinct minima, as displayed in Fig. \ref{He3_Dens} (b).

For comparison, we have plotted the density profiles for the
initial 1s state at $E_0=30$ a.u. and  $E_0=40$ a.u.
in Fig. \ref{Density_Profiles} (c) and (d).
In contrast to the case of the 2s state,
for both peak field strengths the distinct nodal structure of the wave function
near the nucleus is absent at the beginning of the turn-off of the laser pulse,
such that the photoionization cross section is not reduced during the turn-off
of the pulse.
Consequently, the overall ionization probabilities do not exhibit
a further decrease in addition to that caused by the stabilization effect.
Furthermore, we have performed calculations employing the initial 3s state,
which has an ionization potential of $E_0=0.436$ a.u.,
and exhibits two nodes in the wave function density at $r=1.05$ a.u. and
$r=3.28$ a.u., respectively.
For this particular choice of the initial state, as with the 1s state
we do not observe any ionization suppression.
Invoking the analogous transfer mechanism as in the 2s case,
we have calculated according to Eq. (\ref{Transition_Operator}) the
corresponding function,
$ P_{13}=\left| \left< \Phi_{\rm KH} | \Phi_{\rm 1s}\right>
    \left< \Phi_{\rm 3s} | \Phi_{\rm KH}\right> \right|^2$ .
This function, however, is more than an order of magnitude smaller than that
of the 2s state,
and we estimate that it exhibits a maximum at field strengths higher than
those considered here,
such that the ionization suppression effect can not be expected to occur
in the 3s state.

Finally, we stress the role of the retardation of the laser pulse
by comparing the calculated ionization probabilities with results
obtained within the dipole approximation.
The latter have been computed by explicitly neglecting the spatial
dependence of the vector potential,
and therefore also that of the electric field.
At the high laser intensities and frequencies employed
throughout our calculations,
non-dipole effects are expected to have a non-negligible impact
on the atomic ionization dynamics,
and may also suppress the stabilization effect \cite{Gajda_92}.
Comparison of results with and without dipole approximation
for the case of a pulse of nine cycles duration shows
that indeed the stabilization effect at $E_0 \approx 20$ a.u. is more pronounced
in the calculations within the dipole approximation.
At field strengths higher than $E_0=25$ a.u., the non-dipole results
for the ionization probabilities start to exceed noticeably those calculated
within the dipole approximation.
For the highest field strength $E_0=54$ a.u. the Lorentz force acting on
the electron increases the ionization probability by about forty per cent,
whilst it affects neither the general shape of the ionization probabilities
nor the ionization suppression effect.

In conclusion, we have employed a two-dimensional model incorporating
non-dipole effects of the laser pulse to study the ionization and stabilization
of excited atomic states.
It has been found that beyond the stabilization regime ionization suppression
may occur.
The existence of this effect has been shown to be caused by a strong coupling
of the initial, field-free state to the ground state via the eigenstates
of the laser-driven system,
which can also be understood in terms of the wave function evolution
in time.
Furthermore, it has been found that both stabilization and this effect 
are diminished yet not completely destroyed by the Lorentz force acting
on the electronic wave packet.

Funding by the German Research Foundation, Grant No.~KE 721/1-1,
is gratefully acknowledged.

\end{document}